\documentclass[aps,prl,twocolumn,a4paper]{revtex4-1}
\usepackage[dvipdfmx]{graphicx}

\usepackage{amsmath,amsthm,amssymb}
\usepackage{amsmath,bm}
\usepackage{color}
\usepackage{ifthen}

\usepackage{multirow,bigdelim}

\newcount \commentout
\newcommand{\1}{\mbox{1}\hspace{-0.25em}\mbox{l}}

\newlength{\figwidth}
\newlength{\figlarge}
\begin{document}
\title{
Non-Hermitian topology in rock-paper-scissors games
}

\author{
Tsuneya Yoshida
}
\email[]{yoshida@rhodia.ph.tsukuba.ac.jp}
\author{
Tomonari Mizoguchi
}
\author{
Yasuhiro Hatsugai
}
\affiliation{Department of Physics, University of Tsukuba, Ibaraki 305-8571, Japan}
\date{\today}
\begin{abstract}
{\bf Abstract} Non-Hermitian topology is a recent hot topic in condensed matters. 
In this paper, we propose a novel platform drawing interdisciplinary attention: rock-paper-scissors (RPS) cycles described by the evolutionary game theory. 
Specifically, we demonstrate the emergence of an exceptional point and a skin effect by analyzing topological properties of their payoff matrix. 
Furthermore, we discover striking dynamical properties in an RPS chain: the directive propagation of the population density in the bulk and the enhancement of the population density only around the right edge.
Our results open new avenues of the non-Hermitian topology and the evolutionary game theory.
\end{abstract}
\maketitle

%
%
%
\noindent
{\bf Introduction}\\
%
In these decades, the notion of the topology has played a central role in condensed matter physics~\cite{Kane_Z2TI_PRL05_1,HgTe_Bernevig06,Konig_QSHE2007,TI_review_Hasan10,TI_review_Qi10}. 
Analysis of topological aspects of condensed matters dates back to integer quantum Hall systems~\cite{Klitzing_IQHE_PRL80} 
which exhibit robust chiral edge modes induced by the topology of the Hermitian Hamiltonian in the bulk~\cite{Thouless_PRL1982,Halperin_PRB82,Hatsugai_PRL93}. 
While these robust edge modes are originally reported for quantum systems, they have been extended to various disciplines of science~\cite{Haldane_chiralPHC_PRL08,Wang_chiralPHC_Nature09,Fu_chiralPHCapp_AppPhys10,Ozawa_TopoPhoto_RMP19,ProdanPRL09,Kane_NatPhys13,Kariyado_SR15,Susstrunk_TopoMech_Sci15,Delplace_topoEq_Science17,ActiveMatter_SonePRL19,Yoshida_topodiff_SciRep20}. 
In particular, the notion of topology has been extended to an interdisciplinary field~\cite{Knebel_RPSchain_PRL20,Yoshida_chiralRPS_PRE21}; the emergence of topological edge modes has been predicted~\cite{Yoshida_chiralRPS_PRE21} for networks of rock-paper-scissors (RPS) cycles which are described by the evolutionary game theory~\cite{Weibull_textbook97,Sigmund_Game_AMS11,Kirkup_RPSbac_Nat04,Rosas_Social_JTB10,Wang_SocSci_PRE13,Perc_Game_PhysRep17}.

Among the variety of topological phenomena, non-Hermitian topology~\cite{Hatano_PRL96,CMBender_PRL98,Gong_class_PRX18,Kawabata_gapped_PRX19,Bergholtz_Review19,Ashida_nHReview_arXiv20} has become one of the recent hot topics because it results in novel phenomena which do not have Hermitian counterparts. 
A representative example is the emergence of exceptional points (EPs)~\cite{Rotter_EP_JPA09,HShen2017_non-Hermi,VKozii_nH_arXiv17,Yoshida_EP_DMFT_PRB18} (and their symmetry-protected variants~\cite{Budich_SPERs_PRB19,Okugawa_SPERs_PRB19,Yoshida_SPERs_PRB19,Zhou_SPERs_Optica19,Kimura_SPERs_PRB19,Yoshida_SPERs_mech19,Delplace_symmEP3_arXiv21}) on which eigenvalues of the non-Hermitian Hamiltonian touch both for the real- and imaginary-parts. 
Another typical example is a non-Hermitian skin effect~\cite{SYao_nHSkin-1D_PRL18,Lee_Skin19,Zhang_BECskin19,Okuma_BECskin19,JYLee_nHDyn_PRL19,Borgnia_ptGapPRL2020,Yoshida_MSkinPRR20} which is extreme sensitivity of the eigenvalues and eigenstates to the presence/absence of boundaries.
As is the case of the Hermitian topology, the above non-Hermitian phenomena have also been reported in a variety of systems~\cite{Guo_nHExp_PRL09,Zhen_AcciEP_Nat15,Hassan_EP_PRL17,Zhou_ObEP_Science18,Zyuzin_nHEP_PRB18,Papaji_nHEP_PRB19,HShen2018quantum_osci,Matsushita_ER_PRB19,Hofmann_ExpRecipSkin_19,Helbig_ExpSkin_19} (e.g., photonic systems~\cite{Guo_nHExp_PRL09,Zhen_AcciEP_Nat15,Hassan_EP_PRL17,Zhou_ObEP_Science18} and quantum systems~\cite{Zyuzin_nHEP_PRB18,Papaji_nHEP_PRB19,HShen2018quantum_osci,Matsushita_ER_PRB19,Matsushita_nHResp_JPSJ21}).

Despite the above significant progress, topological phenomena of the evolutionary game theory, which attract interdisciplinary attention, are restricted to the Hermitian topology. Highlighting non-Hermitian topology of such systems is significant as it may provide a new insights and may open a new avenue of the evolutionary game theory.

In this paper, we report non-Hermitian topological phenomena in the evolutionary game theory: an EP and a skin effect in RPS cycles. The EP in the single RPS cycle is protected by the realness of the payoffs which is mathematically equivalent to parity-time ($PT$) symmetry. Our linearized replicator equation elucidates that the EP governs dynamics of the RPS cycle. Furthermore, we discover striking dynamical phenomena in an RPS chain induced by the skin effect: the directive propagation of the population density in the bulk and the enhancement of the population density only around the right edge.
These dynamical properties are in sharp contrast to those in the Hermitian systems. The above results open new avenues of the non-Hermitian topology and the evolutionary game theory.
%
%
\\
\\
\noindent
{\bf Results}\\
\noindent
{\bf EP in a single RPS cycle.}
%
Firstly, we demonstrate the emergence of an EP at which two eigenvalues touch both for the real- and imaginary-parts.

\begin{figure}[!h]
\begin{minipage}{1\hsize}
\begin{center}
\includegraphics[width=1\hsize,clip]{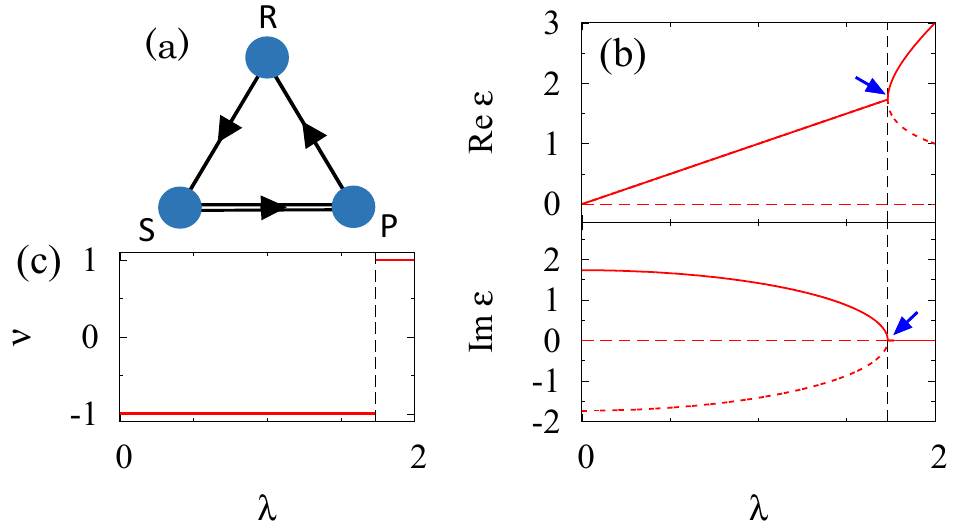}
\end{center}
\end{minipage}
\caption{
(a): Sketch of the RPS cycle. The arrows denote dominance relationship between the strategies for $\lambda=0$; ``R" beats ``S"; ``S" beats ``P"; ``P" beats ``R".
The second terms proportional to $\lambda$ are introduced between sites connected by the double line.
(b): The real- and imaginary-parts of eigenvalues as functions of $\lambda$. An eigenvalue of $A(\lambda)$ is zero for an arbitrary $\lambda$ (see horizontal dashed lines).
(c): The $\mathbb{Z}_2$-invariant characterizing the EP.
The dashed vertical lines in panels (b) and (c) denote $\lambda=\lambda_{\mathrm{c}}$.
}
\label{fig: EP spec}
\end{figure}

Consider two players play the RPS game [see Fig.~\ref{fig: EP spec}(a)] whose payoff matrix is given by~\cite{Tainakaa_PLA1993,Juul_PRE2012,Szolnoki_PRE2014,Szolnoki2016}
\begin{eqnarray}
\label{eq: single RPS payoff}
A(\lambda) &=&
\left(
\begin{array}{ccc}
0   & -1 & 1  \\
1   & 0  & -1 \\
-1  & 1  & 0
\end{array}
\right)
+\lambda
\left(
\begin{array}{ccc}
0  & 0  & 0  \\
0  & 1  & -1 \\
0  & -1 & 1
\end{array}
\right),
\end{eqnarray}
with a real number $\lambda$.
Here, players choose one of the strategies $(s_1,s_2,s_3)=(``\mathrm{R}",``\mathrm{P}",``\mathrm{S}")$. 
The payoff of a player is $A_{IJ}$ when the player chooses the strategy $s_I$ and the other player chooses $s_J$ ($I,J=1,2,3$).
For $\lambda=0$, this game is reduced to the standard zero-sum RPS game where the sum of all players' payoff is zero for an arbitrary set of strategies.

The above payoff matrix exhibits an EP, which can be deduced by noting the following two facts.
(i) The first (second) term is anti-Hermitian (Hermitian).
(ii) For an arbitrary $\lambda$, eigenvalues $\epsilon_n$ ($n=1,2,3$) form a pair $\epsilon_n=\epsilon^*_{n'}$ ($n\neq n'$) or are real numbers $\epsilon_n \in \mathbb{R}$.
This constraint arises from the realness of the payoffs
\begin{eqnarray}
\label{eq: PT symm}
\mathcal{K} A(\lambda) \mathcal{K} &=& A(\lambda),
\end{eqnarray}
where the operator $\mathcal{K}$ takes complex conjugate.
Equation~(\ref{eq: PT symm}) can be recognized as $PT$ symmetry by regarding $\lambda$ as a momentum (for more details, see Sec.~S1 of Supplemental Material~{[68]}).
For $\lambda=0$, the energy eigenvalues are aligned along the imaginary axis due to the anti-Hermiticity of $A(\lambda=0)$. 
Increasing $\lambda$, two eigenvalues touch at a critical value $\lambda_\mathrm{c}$ so that they become real numbers when the second term is dominant.
At $\lambda=\lambda_{\mathrm{c}}$, the EP emerges.

The emergence of the EP is supported by Fig.~\ref{fig: EP spec}(b).
For $\lambda=0$, the energy eigenvalues are pure imaginary due to anti-Hermiticity of $A(\lambda=0)$.
As $\lambda$ is turned on, two eigenvalues approach each other, and the EP emerge at $\lambda=\lambda_c=\sqrt{3}$.
One can also characterize the topology of this EP by computing the $\mathbb{Z}_2$-invariant $\nu$~\cite{Delplace_symmEP3_arXiv21}
\begin{eqnarray}
\label{eq: Z2 disc}
\nu(\lambda) &=&\mathrm{sgn}\left[ \mathrm{Disc}_E P(E,\lambda) \right],
\end{eqnarray}
with $P(E,\lambda)=\mathrm{det}[A(\lambda)-E\1]$. Here, $\mathrm{sgn}(x)$ takes $1$ ($-1$) for $x>0$ ($x<0$), and $\mathrm{Disc}_E P(E,\lambda)$ denotes discriminant of $P(E,\lambda)$.
For the $3\times 3$-matrix $A(\lambda)$, $\mathrm{Disc}_E P(E,\lambda)=(\epsilon_1-\epsilon_2)^2(\epsilon_1-\epsilon_3)^2(\epsilon_2-\epsilon_3)^2$ holds.
Figure~\ref{fig: EP spec}(c) plots the $\mathbb{Z}_2$-invariant as a function of $\lambda$.
Corresponding to the emergence of the EP, the $\mathbb{Z}_2$-invariant jumps at $\lambda=\lambda_{\mathrm{c}}$, elucidating the topological protection of the EP.
%
%
\\
\\
{\bf
Dynamical properties and the EP.
}
%
Suppose that a large number of players repeat the above game whose dynamics is described by the replicator equation [see Eq.~(\ref{eq: Rep eq})].
In this case, the EP governs the dynamical behaviors when the population density slightly deviates from a fixed point; the population density shows an oscillatory behavior for $0\leq \lambda<\lambda_{\mathrm{c}}$, while such a behavior disappear for $\lambda_{\mathrm{c}}<\lambda$ due to the emergence of the EP.

Firstly, we linearize the replicator equation.
When a large number of players repeat the game, the time-evolution is described by the replicator equation
\begin{eqnarray}
\label{eq: Rep eq}
\partial_t \bm{e}^T_I \cdot \bm{x} &=& \bm{e}^T_I \cdot \bm{x} \left( \bm{e}^T_I A \bm{x} -\bm{x}^TA\bm{x} \right),
\end{eqnarray}
where $\bm{x}$ denotes the population density ($\sum_I x_I=1$), and $\bm{e}_I$ is a unit vector whose $I$-th element is unity.
The second term vanishes when the payoff matrix $A$ is anti-symmetric.
However, in order to access the non-Hermitian topology, the second term is inevitable which makes the argument in Ref.~\onlinecite{Yoshida_chiralRPS_PRE21} unavailable.

Nevertheless, we can still obtain the following linearized equation
\begin{eqnarray}
\label{eq: linear Rep eq}
\partial_t \delta\bm{x} &=& \frac{1}{N_0} A \delta \bm{x},
\end{eqnarray}
which is mathematically equivalent to the Schr\"odinger equation.
This mathematical equivalence reveals that the dynamics of the RPS cycle (i.e., a classical system) can be understood in terms of quantum physics.
Here, $\delta \bm{x}$ is defined as $\delta \bm{x}=\bm{x}-\bm{c}$ with $\bm{c}=(1,1,\ldots,1)^T/N_0$, and $N_0$ denotes dimensions of the matrix $A$.
Key ingredients are the following relations:
\begin{eqnarray}
\label{eq: sum Aij=0}
\sum_{J}A_{IJ}=0 \quad &\mathrm{and}& \quad \sum_{J}A_{JI}=0,
\end{eqnarray}
for an arbitrary $I$. 
This equation guarantees that $\bm{c}$ denotes a fixed point. 
Linearizing the replicator equation around this fixed point, we obtain Eq.~(\ref{eq: linear Rep eq}) as detailed in the ``Methods".

\begin{figure}[!h]
\begin{minipage}{1\hsize}
\begin{center}
\includegraphics[width=1\hsize,clip]{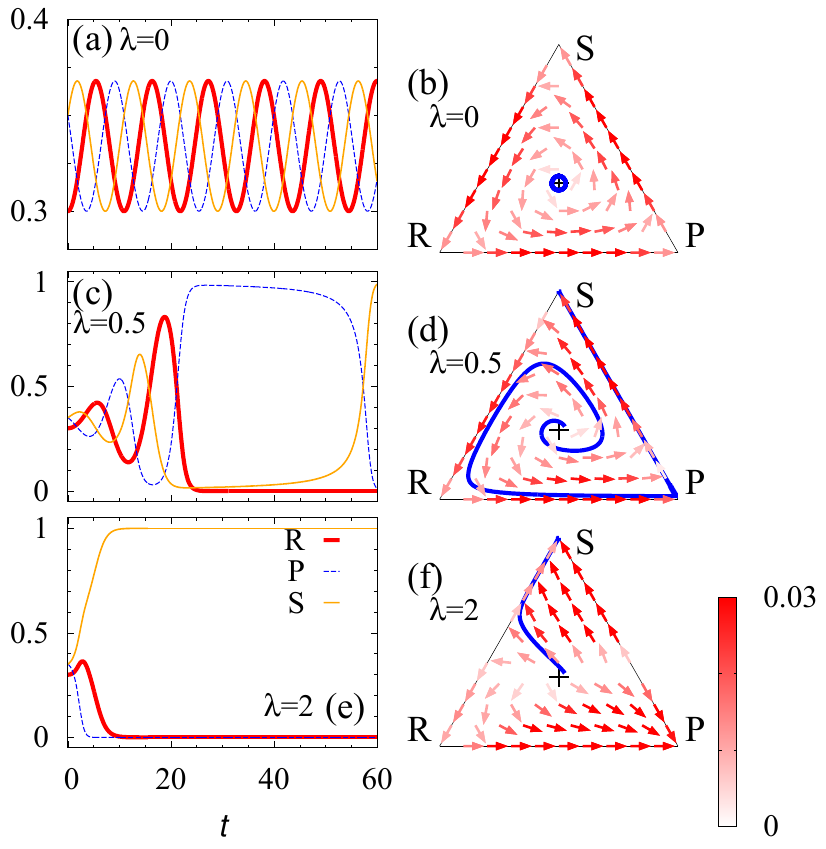}
\end{center}
\end{minipage}
\caption{
Dynamical properties of the RPS game~(\ref{eq: single RPS payoff}).
Panels (a), (c), and (e) display the time-evolution of the population $\bm{x}$ for $\lambda=0$, $0.5$, and $2$, respectively.
The data are obtained by employing a fourth order Runge-Kutta method~\cite{Suli_RK_textbook03} with discretized time $t_n=n\Delta t_{\mathrm{RK}}$ with $\Delta t_{\mathrm{RK}}=0.05$.
We set the initial condition as $\bm{x}(t=0)=(1-\delta_0,1+\delta_0/2,1+\delta_0/2)/3$ with $\delta_0=0.1$. The time-evolution is computed up to $t_n=120$.
Panels (b), (d), and (f) are phase plots for $\bm{x}$. 
The arrows denote the direction of the dynamics $(\Delta x_1,\Delta x_2,\Delta x_3)$ which is computed by
$\Delta x_I=x_I (\bm{e}^T_IA\bm{x}-\bm{x}^TA\bm{x})\Delta t$ with $\Delta t=0.1$.
The black crosses denote the fixed point specified by $\bm{c}=(1,1,1)/3$.
In panels (b), (d), and (f), the blue line denotes the population density $\bm{x}(t)$ plotted in panels (a), (c), and (e), respectively.
}
\label{fig: EP dyn}
\end{figure}

The linearized replicator equation indicates that the dynamics is governed by the spectrum of $A$.
In particular, the imaginary-part of the eigenvalues results in oscillatory behaviors.

To verify the above statement, we numerically solve the replicator equation~(\ref{eq: Rep eq}).
Figure~\ref{fig: EP dyn} plots the time-evolution of the population density for several values of $\lambda$.
For $\lambda=0$, the matrix $A(\lambda)$ is reduced to the payoff matrix of the standard zero-sum RPS game.
In this case, the eigenvalues of $A$ are pure imaginary, which results in the oscillatory behavior [see Fig.~\ref{fig: EP dyn}(a)].
This oscillatory behavior is also observed in the phase plot [see Fig.~\ref{fig: EP dyn}(b)], where the orbit forms a closed loop.
The above oscillatory behavior is also observed for $\lambda=0.5$ [see Figs.~\ref{fig: EP dyn}(c)~and~\ref{fig: EP dyn}(d)], which is due to the imaginary-part of the eigenvalues.
We also note that the deviation $\delta \bm{x}$ is enhanced as $t$ increases, which is because the real-part of the eigenvalues is positive [see Fig.~\ref{fig: EP spec}(b)].
Further increasing $\lambda$ induces the EP [see Fig.~\ref{fig: EP spec}(b)], and the eigenvalues become real.
Correspondingly, the above oscillatory behavior is not observed for $\lambda=2$ [see Figs.~\ref{fig: EP dyn}(e)~and~\ref{fig: EP dyn}(f)],
The above numerical data verify that the EP governs the dynamics around the fixed point $\bm{c}$; the oscillatory behavior of the RPS cycle disappears as the EP emerges.
We note that for $\lambda < 0$, the fixed point corresponds to evolutionary stable strategy. A similar EP emerges also in this case which governs the dynamics.

We close this part with three remarks.
Firstly, so far, we have seen the emergence of the EP in the RPS game by changing $\lambda$.
We note that symmetry-protected exceptional rings are also observed by introducing an additional parameter (see Sec.~{S2} of Supplemental Material~{[68]}), whose topology is also characterized by the $\mathbb{Z}_2$-invariant $\nu$.
Secondly, although Refs.~\onlinecite{Szabo_PhysRep2007,Szolnoki_GameRev_JRSI14,Dobramysl_JPAMT2018,Szolnoki_IT_EPL2020} discusses topology of interaction networks among strategies (i.e., interaction topology), it differs from the topology discussed in this paper.
Thirdly, we note that EPs are also reported for active matters~\cite{Sone_ActiveMatter_NatComm20,Tang_StochasticTopo_PRX21}.
We would like to stress, however, that significance of this paper is to reveal the emergence of the EPs and how they affect the dynamics in systems of the evolutionary game theory which describes the population density of biological systems and human societies.
\\
\\
{\bf 
Skin effect in an RPS chain.
}
%
Now, we discuss a one-dimensional system showing the skin effect whose origin is the non-trivial topology characterized by the winding number~\cite{Zhang_BECskin19,Okuma_BECskin19}.
Because of this non-trivial topology, switching from the periodic boundary condition (PBC) to the open boundary condition (OBC) significantly changes the spectrum. 
Correspondingly, almost of all right eigenstates are localized around the right edge which are called skin modes.

\begin{figure}[!h]
\begin{minipage}{1\hsize}
\begin{center}
\includegraphics[width=1\hsize,clip]{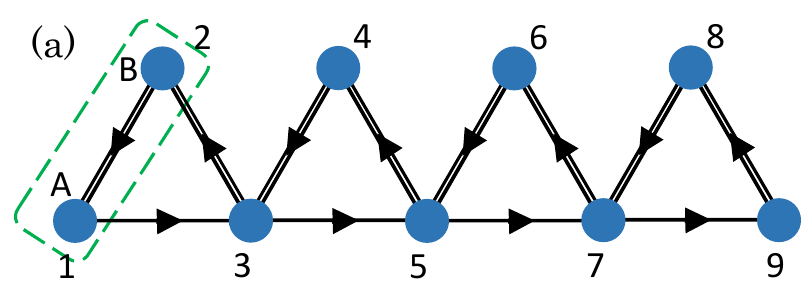}
\end{center}
\end{minipage}
\begin{minipage}{1\hsize}
\begin{center}
\includegraphics[width=1\hsize,clip]{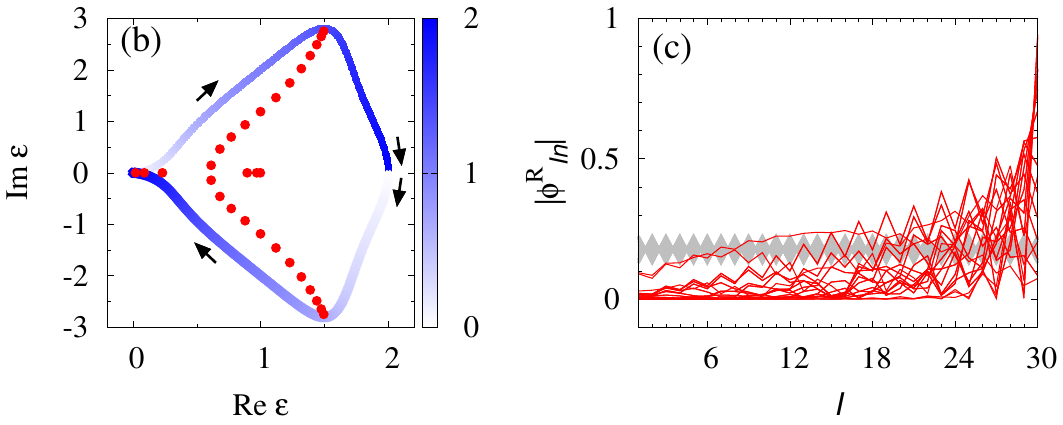}
\end{center}
\end{minipage}
\caption{
(a): Sketch of the RPS chain for $L_x=4$ with an additional site at $I=9$. The arrows describes payoffs. 
The terms proportional to $\lambda$ are introduced between sites connected by double lines.
As shown in this panel, $I$ takes $I=1,2,3,\ldots$. For the PBC, $I+2L_x=I$ holds. Dashed line denotes the unit cell. For $J=1,2,\ldots$, $R_{2I-1}= R_{2I}$ holds. 
(b): Spectrum of the RPS chain for $\lambda=-0.5$. 
The data colored with blue are obtained under the PBC. Here, the shade of color denotes $k/\pi$. The data denoted by red dots are obtained for $L_x=15$ and the OBC. 
(c): Amplitude of the right eigenvectors $\phi^{\mathrm{R}}_{In}$ ($n=1,2,\ldots,2L_x$) as functions of $I$ for $L_x=15$. Red (gray) lines denote data for the OBC (PBC).
}
\label{fig: nHSkin spec}
\end{figure}

The above skin effect can be observed in an RPS chain illustrated in Fig.~\ref{fig: nHSkin spec}(a).
Applying the Fourier transformation, the payoff matrix is written as
\begin{eqnarray}
A(k)&=& 
\left(
\begin{array}{cc}
0 & 1-e^{-ik} \\
-1+e^{ik} & -2i\sin k
\end{array}
\right)
+
\lambda
\left(
\begin{array}{cc}
-2 & 1+e^{-ik} \\
 1+e^{ik} & -2
\end{array}
\right),\nonumber \\
\end{eqnarray}
under the PBC.
Here, $\bm{x}_{k}$ is defined as $\bm{x}^T_{k}=(x_{k\mathrm{A}},x_{k\mathrm{B}})$ with  $x_{k\alpha}=\frac{1}{L_x} \sum_{R_I} e^{ik R_I} x_{R_I\alpha} $ and $\alpha=\mathrm{A},\mathrm{B}$.
Sets of $R_I$ and $\alpha_{I}$ are specified by $I$ ($x_I=x_{R_I\alpha_I}$).
For the explicit form of the payoff matrix in the real-space, see Sec.~{S3} of Supplemental Material~{[68]}.

Figure~\ref{fig: nHSkin spec}(b) plots the spectrum of the payoff matrix for $\lambda=-0.5$.
When the PBC is imposed, eigenvalues form a loop structure as denoted by blue lines in Fig.~\ref{fig: nHSkin spec}(b).
Accordingly, the winding number~\cite{Gong_class_PRX18,Zhang_BECskin19,Okuma_BECskin19} defined as
\begin{eqnarray}
\label{eq: winding}
W &=& \oint \frac{dk}{2\pi i} \partial_k \log \mathrm{det}[A(k)-\epsilon_{\mathrm{ref}}\1 ],
\end{eqnarray}
takes $-1$ for $\epsilon_{\mathrm{ref}}=1$, which implies the skin effect.
Indeed, imposing the OBC significantly changes the spectrum [see red dots in Fig.~\ref{fig: nHSkin spec}(b)].
Correspondingly, almost all of the right eigenvectors are localized around the edges, meaning the emergence of skin modes [see Fig.~\ref{fig: nHSkin spec}(c)].
%
%
\begin{figure}[!h]
\begin{minipage}{1\hsize}
\begin{center}
\includegraphics[width=1\hsize,clip]{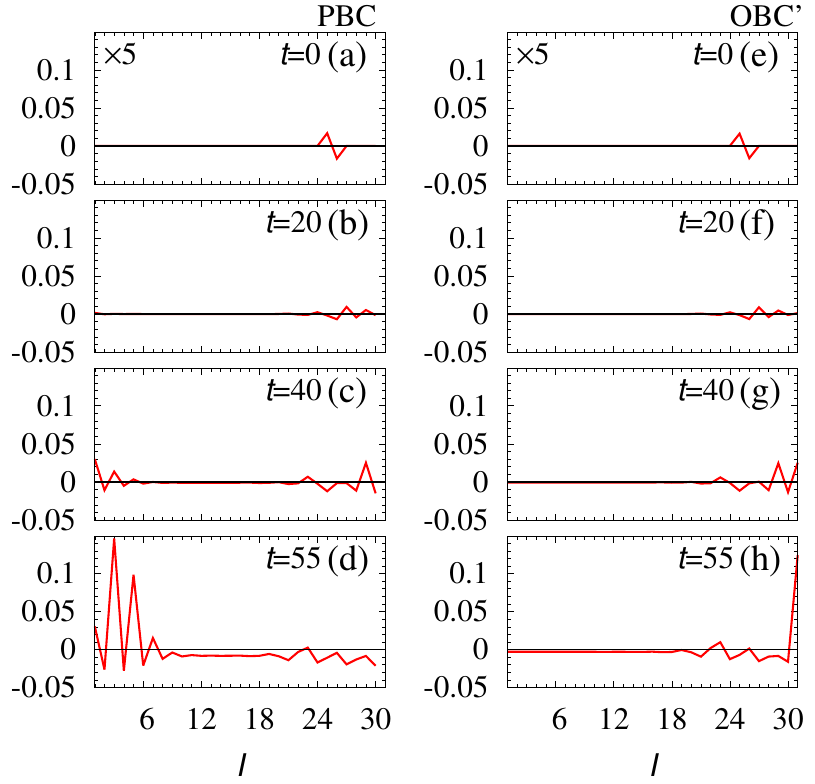}
\end{center}
\end{minipage}
\caption{
Time-evolution of the population density $\delta \bm{x}(t)=\bm{x}(t)-\bm{c}$ for $\lambda=-0.5$ and $L_x=15$.
The horizontal axis denotes $I$. (a)-(d) [(e)-(f)]: The time-evolution under the PBC [OBC']. 
The data in panels (a) and (e) are multiplied by 5. 
The data are obtained by employing the fourth order Runge-Kutta method~\cite{Suli_RK_textbook03} with discretized time $t_n=n\Delta t_{\mathrm{RK}}$ with $\Delta t_{\mathrm{RK}}=0.05$. 
We set the initial condition so that $N_0 x_{I}(0)$ takes $1+\delta_0$ and $1-\delta_0$ with $\delta_0=0.1$ for $I=25$ and $I=26$, respectively. 
For the other $I$, $N_0 x_{I}(0)$ takes 1. 
Here, $N_0$ is chosen so that $\sum_Ix_I(0)=1$ holds.
}
\label{fig: nHKSkin dyn}
\end{figure}
%
The above data [Figs.~\ref{fig: nHSkin spec}(b)~and~\ref{fig: nHSkin spec}(c)] indicate that the skin effect is observed in the RPS chain.
Here, we note that the spectrum and skin modes are almost unchanged under the following perturbation to the edges (see Sec.~{S3} of Supplemental Material~{[68]}): attaching site at  $I=2L_x+1$ and tuning diagonal elements to $A_{II}=\lambda$ only for $I=1,2L_x+1$ so that Eq.~(\ref{eq: sum Aij=0}) holds [see Fig.~\ref{fig: nHSkin spec}(a)]. We refer to this boundary condition as OBC'.

The above skin effect results in striking dynamical properties: the directive propagation of the population density in the bulk and the enhancement of the population density only around the right edge.
The directive propagation is observed by imposing the PBC. 
Figures~\ref{fig: nHKSkin dyn}(a)-\ref{fig: nHKSkin dyn}(d) indicate that the propagation only to the right direction is enhanced. 
This phenomenon can be understood by noting the following facts as well as linearized approximation: (i) as $t$ increases, each mode is enhanced corresponding to $\mathrm{Re}\epsilon_n$; (ii) the group velocity of each mode is proportional to $-\partial_k \mathrm{Im} \epsilon_n$. 
Because of the loop structure resulting in $W=-1$, the modes propagating to the right are more enhanced than the ones propagating to the left. 

The enhancement of the population density at the right edge is observed by imposing OBC'. 
We note that Eq.~(\ref{eq: sum Aij=0}) is satisfied for OBC' while it is not for OBC.
As shown in Figs.~\ref{fig: nHKSkin dyn}(e)-\ref{fig: nHKSkin dyn}(h), the population density around the right edge is enhanced as $t$ increases. 
This phenomenon is due to the skin modes whose eigenvalues satisfy $\mathrm{Re}\epsilon_n>0$.
The above dynamical behaviors are unique to non-Hermitian systems; turning off $\lambda$, $iA(\lambda=0)$ becomes Hermitian and the above behaviors disappear (see Sec.~{S3} of Supplemental Material~{[68]}).

We close this part with two remarks.
Firstly, we note that a similar behaviors can be observed for another RPS chain, implying ubiquity of the skin effect (see Sec.~{S4} of Supplemental Material~{[68]}).
Secondly, Ref.~\onlinecite{Knebel_RPSchain_PRL20} has analyzed an RPS chain whose payoff matrix is Hermitian up to the imaginary unit $i$. 
We would like to stress, however, that our aim is to observe non-Hermitian topological phenomena which are not accessible with the model in Ref.~\onlinecite{Knebel_RPSchain_PRL20}.
In this non-Hermitian case, we need to take into account the second term of Eq.~(\ref{eq: Rep eq}).
%
\\
\\
{\bf 
Discussion
}
We have proposed a new platform of non-Hermitian topology in an interdisciplinary field, i.e., RPS cycles described by the evolutionary game theory.
Specifically, by analyzing the payoff matrix, we have demonstrated the emergence of the EP and the skin effect which are representative non-Hermitian topological phenomena. 
In addition, our linearized replicator equation has revealed that the EP governs the dynamics of the population density around the fixed point. 
Furthermore, we have discovered the striking dynamical phenomenon in the RPS chain: the directive propagation of the population density in the bulk and the enhancement of the population density only around the right edge which are induced by the skin effect. 

Our results pose several future directions which we discuss below.
The experimental observation is one of the central issues. 
In particular, our results provide a first step toward the observation of the non-Hermitian topology beyond natural science because the game theory describes a wide variety of systems from biological systems~\cite{Kirkup_RPSbac_Nat04,Sinervo_RPSLizard_Nature96} to human societies~\cite{Rosas_Social_JTB10,Szolnoki_PRL12,Szolnoki_PRX13,Wang_SocSci_PRE13,Perc_Game_PhysRep17}; for instance, dynamics of human cooperation has been discussed in Refs.~\onlinecite{Szolnoki_PRL12,Szolnoki_PRX13}. 
The experimental observation in such a system is considered to be a significant step to the application of topological phenomena beyond natural science. 
In addition, as is the case of equatorial wave~\cite{Delplace_topoEq_Science17}, our result may provide a novel perspective of well-known phenomena for biological systems such as bacteria~\cite{Kirkup_RPSbac_Nat04} and side-blotched lizards~\cite{Sinervo_RPSLizard_Nature96}; dynamics of such systems may be understood in terms of exceptional points.
We also note that the topological classification for systems described by the game theory is also a crucial issue to be addressed.
\\
\\
\noindent
{\bf Methods}
\\
\noindent
{\bf Derivation of the linearized replicator equation.}
Here, we derive Eq.~(\ref{eq: linear Rep eq}).
We start with noting Eq.~(\ref{eq: sum Aij=0}) results in the relations $A\bm{c}=0$ and $\bm{c}^TA=0$ which mean that $\bm{c}=(1,1,1,\ldots,1)^T/N_0$ is a fixed point.
By making use of the above relations we have
\begin{eqnarray}
&&\partial_t \bm{e}^T_I \cdot \delta \bm{x} \nonumber \\
&&= \bm{e}^T_I \cdot (\bm{c}+\delta\bm{x}) \left[ \bm{e}^T_I A (\bm{c}+\delta \bm{x}) -(\bm{c}+\delta \bm{x})^T A (\bm{c}+\delta \bm{x}) \right] \nonumber \\
&&= \bm{e}^T_I \cdot (\bm{c}+\delta\bm{x}) \left( \bm{e}^T_I A \delta \bm{x} -\delta \bm{x}^TA \delta \bm{x} \right) \nonumber \\
&&\sim (\bm{e}^T_I \cdot \bm{c}) \bm{e}^T_I A \delta \bm{x}. 
\end{eqnarray}
From the second to the third line, we have used the relations $A\bm{c}=0$ and $\bm{c}A=0$.
In the last line, we have discarded the second and third order terms of $\delta\bm{x}$.

Because $(\bm{e}^T_I \cdot \bm{c})=1/N_0$ for an arbitrary $I$, we have
\begin{eqnarray}
\partial_t \delta \bm{e}^T_I\cdot \bm{x}&\sim& \frac{1}{N_0} \bm{e}^T_I A \delta \bm{x},
\end{eqnarray}
which is equivalent to Eq.~(\ref{eq: linear Rep eq}).



%
{\bf Acknowledgements.}
This work is supported by JSPS Grant-in-Aid for Scientific Research on Innovative Areas ``Discrete Geometric Analysis for Materials Design": Grant No.~JP20H04627.
This work is also supported by JSPS KAKENHI Grants No.~JP17H06138, No.~JP20K14371, and No.~JP21K13850.
\\
\\
{\bf Author contributions}
T.Y., T. M., and Y. H. planned the project. T. Y. performed the calculations. T.Y., and T. M., and Y. H. were involved in the discussion of the materials and the preparation of the manuscript.
\\
\\
{\bf Competing financial interests}
The authors declare no competing interests.

\clearpage

\renewcommand{\thesection}{S\arabic{section}}
\renewcommand{\theequation}{S\arabic{equation}}
\setcounter{equation}{0}
\renewcommand{\thefigure}{S\arabic{figure}}
\setcounter{figure}{0}
\renewcommand{\thetable}{S\arabic{table}}
\setcounter{table}{0}
\makeatletter
\c@secnumdepth = 2
\makeatother

\onecolumngrid
\begin{center}
 {\large \textmd{Supplemental Materials:} \\[0.3em]
 {\bfseries 
 Non-Hermitian topology in rock-paper-scissors games
 }
 }
\end{center}

\setcounter{page}{1}


\section{
$PT$ symmetry
}
\label{sec: PT symm app}

Here, we briefly review topological properties under $PT$ symmetry (i.e., symmetry described by the product of the time-reversal operation and inversion).
When the system is $PT$ symmetric, the Bloch Hamiltonian $h(\bm{k})$ satisfies
\begin{eqnarray}
\label{eq: PT symm app}
PT h(\bm{k}) PT^{-1} &=& h(\bm{k}),
\end{eqnarray}
with an anti-unitary operator $PT=U_{PT}\mathcal{K}$, a unitary matrix $U_{PT}$, and $\bm{k}$ denoting the momentum.
Here, each of the time-reversal operation and the spatial inversion flips the momentum ($\bm{k}\to -\bm{k}$), and thus, their product does not flip $\bm{k}$.

Equation~(\ref{eq: PT symm app}) results in the following constraint on the eigenvalue $\epsilon_n$ ($n=1,2,\ldots$)
\begin{eqnarray}
\label{eq: e=e^* or e in R app}
\epsilon_n=\epsilon^*_{n'} \ &\mathrm{or}& \ \epsilon_n \in \mathbb{R}.
\end{eqnarray}
The above equation is obtained by a straightforward calculation.
Suppose that $|\psi_{n}(\bm{k}) \rangle_R$ and $\epsilon_n(\bm{k})$ are a right eigenvector and an eigenvalue of $h(\bm{k})$, $h(\bm{k})|\psi_{n}(\bm{k}) \rangle_R=\epsilon_n(\bm{k})|\psi_{n}(\bm{k}) \rangle_R$.
Then, $PT |\psi_{n}(\bm{k}) \rangle_R$ is an eigenstate with eigenvalue $\epsilon^*_n$ because
\begin{eqnarray}
h(\bm{k}) PT |\psi_{n}(\bm{k}) \rangle_R&=& PT  h(\bm{k})  |\psi_{n}(\bm{k}) \rangle_R \nonumber \\
                                                       &=& \epsilon^*_n(\bm{k}) PT  | \psi_{n}(\bm{k}) \rangle_R,
\end{eqnarray}
holds. 
Therefore, we obtain Eq.~(\ref{eq: e=e^* or e in R app}).

Now, let us discuss the topology for $PT^2=1$.
When a point-gap opens at $\epsilon_{\mathrm{ref}} \in \mathbb{R}$ [i.e., $\mathrm{det}(h-\epsilon_{\mathrm{ref}})\neq 0$ holds], the system may possess topological properties which are characterized by the following zero-dimensional $\mathbb{Z}_2$-invariant $\nu'$
\begin{eqnarray}
\label{eq: Z2 det app}
\nu' &=& -\mathrm{sgn}[\mathrm{det}(h-\epsilon_{\mathrm{ref}})],
\end{eqnarray}
with $\mathrm{sgn}(x)$ takes $1$ ($-1$) for $x>0$ ($x<0$).
For $\epsilon_{\mathrm{ref}}=\epsilon_{\mathrm{EP}}$ ($\epsilon_{\mathrm{EP}}=\sqrt{3}$), $\nu'$ and $\nu$ defined in Eq.~(3) characterizes the same topology of Eq.~(1);
because $\nu'=\prod_n (\epsilon_n-\epsilon_\mathrm{\mathrm{ref}})$ holds, we have $\nu'$ taking $1$ ($-1$) for $\lambda>\lambda_{\mathrm{c}}$ ($\lambda<\lambda_{\mathrm{c}}$).

\section{
Symmetry-protected exceptional rings in an RPS cycle
}
\label{sec: SPER app}
We demonstrate the emergence of a symmetry-protected exceptional rings in evolutionary game theory.
Consider an extended RPS cycle whose payoff matrix is written as
\begin{eqnarray}
A(\lambda,\kappa)&=& 
A(\lambda)+\kappa
\left(
\begin{array}{ccc}
1  & 0  & -1 \\
0  & 0  & 0 \\
-1 & 0  & 1
\end{array}
\right),
\end{eqnarray}
with $\lambda,\kappa\in \mathbb{R}$.

We can block-diagonalize the matrix $A$.
Applying the unitary transformation with 
\begin{eqnarray}
V&=& 
\left(
\begin{array}{ccc}
\frac{1}{\sqrt{3}} &  \frac{1}{\sqrt{2}} & \frac{1}{\sqrt{6}} \\
\frac{1}{\sqrt{3}} & -\frac{1}{\sqrt{2}} & \frac{1}{\sqrt{6}} \\
\frac{1}{\sqrt{3}} &         0           & \frac{-\sqrt{2}}{\sqrt{3}}
\end{array}
\right),
\end{eqnarray}
we have
\begin{eqnarray}
V^TA(\lambda, \kappa )V &=& 
\left(
\begin{array}{ccc}
0 &  0 &  0 \\
0 & 0  &  -\sqrt{3} \\
0 & \sqrt{3}   & 0
\end{array}
\right)
+\lambda
\left(
\begin{array}{ccc}
0  & 0 &  0 \\
0  & \frac{1}{2}& -\frac{\sqrt{3}}{2}  \\
0  & -\frac{\sqrt{3}}{2} & \frac{3}{2}
\end{array}
\right)
+\kappa
\left(
\begin{array}{ccc}
0  & 0 &  0 \\
0  & \frac{1}{2}& \frac{\sqrt{3}}{2}  \\
0  & \frac{\sqrt{3}}{2} & \frac{3}{2}
\end{array}
\right).
\end{eqnarray}

Diagonalizing the above matrix, we have eigenvalues $0$ and 
\begin{eqnarray}
\epsilon_\pm&=& (\lambda+\kappa) \pm \sqrt{\frac{3}{4}(\lambda-\kappa)^2+\frac{1}{4}(\lambda+\kappa)^2-3}.
\end{eqnarray}
We note that for $\kappa=0$, the EP emerges at $\lambda=\lambda_{\mathrm{c}}= \sqrt{3}$.

\begin{figure}[!h]
\begin{minipage}{1\hsize}
\begin{center}
\includegraphics[width=1\hsize,clip]{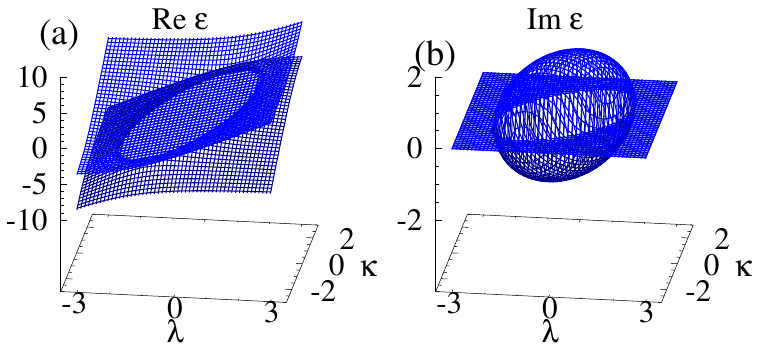}
\end{center}
\end{minipage}
\caption{
(a) and (b): The real- and imaginary-parts of $\epsilon_{\pm}$. The data colored with blue (dark blue) represents $\epsilon_+$ ($\epsilon_-$).
}
\label{fig: SPER app}
\end{figure}

Figure~\ref{fig: SPER app}(a) [\ref{fig: SPER app}(b)] plots the real- [imaginary] part of $\epsilon_\pm$.
These figures indicate the emergence of the symmetry-protected exceptional rings; EPs, where the two bands touch both for the real- and the imaginary-parts, form a ring.
We also note that $\nu$ takes $-1$ ($1$) inside (outside) of the ring, elucidating the topological protection.

The above results verify the emergence of the symmetry-protected exceptional rings in the extended RPS cycle.

\section{
Details of the RPS chain
}
\label{sec: RPS chain app}

We discuss details of the RPS chain defined in the main text.
Under the OBC', the payoff matrix is written as 
\begin{eqnarray}
A(\lambda) &=&A_0 +\lambda B.
\end{eqnarray}
We have introduced an additional site at $I=9$ [see Fig.~3(a)] for $L_x=4$.
Hence, the vector $\bm{x}$ consists of nine components, $\bm{x}=(x_1,x_2,x_3,\cdots,x_{9})^T$.
Here, $A_0$ and $B$ are defined as
\begin{subequations}
\begin{eqnarray}
A_0&=& 
\left(
\begin{array}{ccccc}
A_{0\mathrm{d}}  & A^T_{0\mathrm{c}}   & 0                   & 0                          & 0 \\
A_{0\mathrm{c}}  & A_{0\mathrm{d}}     & A^T_{0\mathrm{c}}   & 0                          & 0  \\
0                & A_{0\mathrm{c}}     & A_{0\mathrm{d}}     & A^T_{0\mathrm{c}}          & 0  \\
0                &  0                  & A_{0\mathrm{c}}     & A_{0\mathrm{d}}            & A^T_{0\mathrm{cR}}  \\
0                &  0                  & 0                   & A_{0\mathrm{cR}}           & 0
\end{array}
\right),
\end{eqnarray}
with
\begin{eqnarray}
A_{0\mathrm{d}}= 
\left(
\begin{array}{cc}
0  &  1\\
-1 &  0
\end{array}
\right),
\quad
A_{0\mathrm{c}}= 
\left(
\begin{array}{cc}
-1  &  0\\
 1  &  0
\end{array}
\right),
\quad
A_{0\mathrm{cR}}
=
\left(
\begin{array}{cc}
1 & -1
\end{array}
\right),
\end{eqnarray}
\end{subequations}
and
\begin{subequations}
\begin{eqnarray}
B&=& 
\left(
\begin{array}{ccccc}
B_{\mathrm{dL}}  & B^T_{\mathrm{c}}   & 0                   & 0                            & 0 \\
B_{\mathrm{c}}   & B_{\mathrm{d}}     & B^T_{\mathrm{c}}   & 0                            & 0  \\
0                 & B_{\mathrm{c}}     & B_{\mathrm{d}}     & B^T_{\mathrm{c}}            & 0  \\
0                 &  0                  & B_{\mathrm{c}}     & B^T_{\mathrm{d}}            & B^T_{\mathrm{cR}}  \\
0                 &  0                  & 0                   & B^T_{\mathrm{cR}}           & -1
\end{array}
\right),
\end{eqnarray}
with
\begin{eqnarray}
B_{\mathrm{d}}= 
\left(
\begin{array}{cc}
-2  &  1\\
1 &  -2
\end{array}
\right),
\quad
B_{\mathrm{c}}= 
\left(
\begin{array}{cc}
0   &  1\\
0   &  0
\end{array}
\right),
\quad
B_{\mathrm{dL}}= 
\left(
\begin{array}{cc}
-1  &  1\\
1 &  -2
\end{array}
\right),
\quad
B_{\mathrm{cR}}
=
\left(
\begin{array}{cc}
0 & 1
\end{array}
\right),
\end{eqnarray}
\end{subequations}
respectively.
In addition, we have imposed $B_{11}=B_{99}=-1$ to satisfy Eq.~(6).

\begin{figure}[!h]
\begin{minipage}{1\hsize}
\begin{center}
\includegraphics[width=1\hsize,clip]{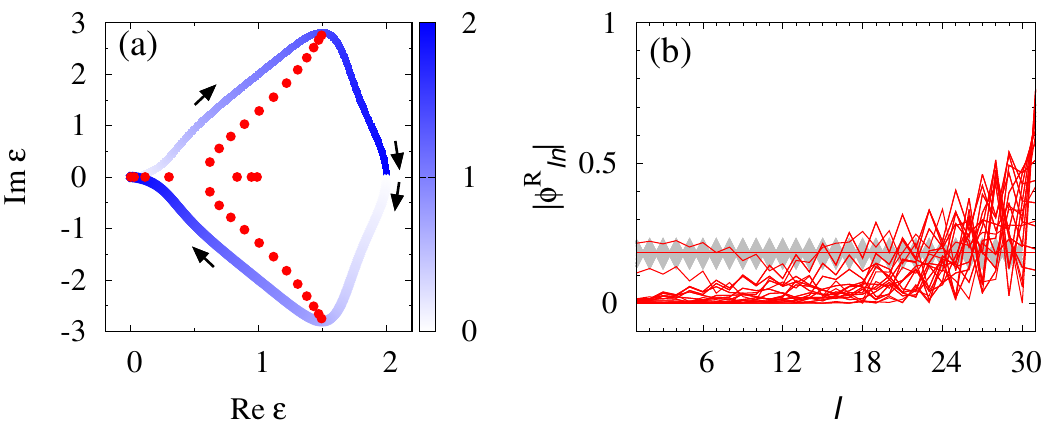}
\end{center}
\end{minipage}
\caption{
Spectrum and amplitude of right eigenvectors under the OBC'.
(a): The spectrum for $\lambda=-0.5$ and $L_x=15$. 
(b):Amplitude of the right eigenvectors $\phi^{\mathrm{R}}_{In}$ ($n=1,2,\ldots,2L_x+1$) as functions of $I$ for $L_x=15$.
The data are plotted in a similar way to Fig.~3.
}
\label{fig: nHSkin OBC' spec app}
\end{figure}

In Fig.~\ref{fig: nHSkin OBC' spec app}, we can see that the spectrum and amplitude of the right eigenvectors for OBC' are similar to those for OBC, which indicates the robustness of the skin effect against perturbations. This is because the non-Hermitian topology induces the skin effect.
We also note that $\bm{c}=(1,1,\ldots,1)/31$ is the right eigenvector for $L_x=15$ [see Fig.~\ref{fig: nHSkin OBC' spec app}(b)] because Eq.~(6) holds under OBC'.

When Eq.~(6) is satisfied, the replicator equation~(4) can be linearized as discussed in the main text.
In this case, the dynamics is governed by
\begin{eqnarray}
i \partial_t \delta \bm{x}(t)&=& H \delta \bm{x}(t),
\end{eqnarray}
with $H=iA/N_0$ [see also Eq.~(5)] which is mathematically equivalent to the Schr\"odinger equation.
In quantum systems, the group velocity and the lifetime are written as $\partial_k \mathrm{Re} E_n(k)$ and $1/\mathrm{Im} E_n(k)$ ($\mathrm{Im} E_n(k)>0$), respectively.
Here, $E_n(k)$ $(n=1,2,\ldots)$ denotes eigenvalues of $H$.

\begin{figure}[!h]
\begin{minipage}{0.6\hsize}
\begin{center}
\includegraphics[width=1\hsize,clip]{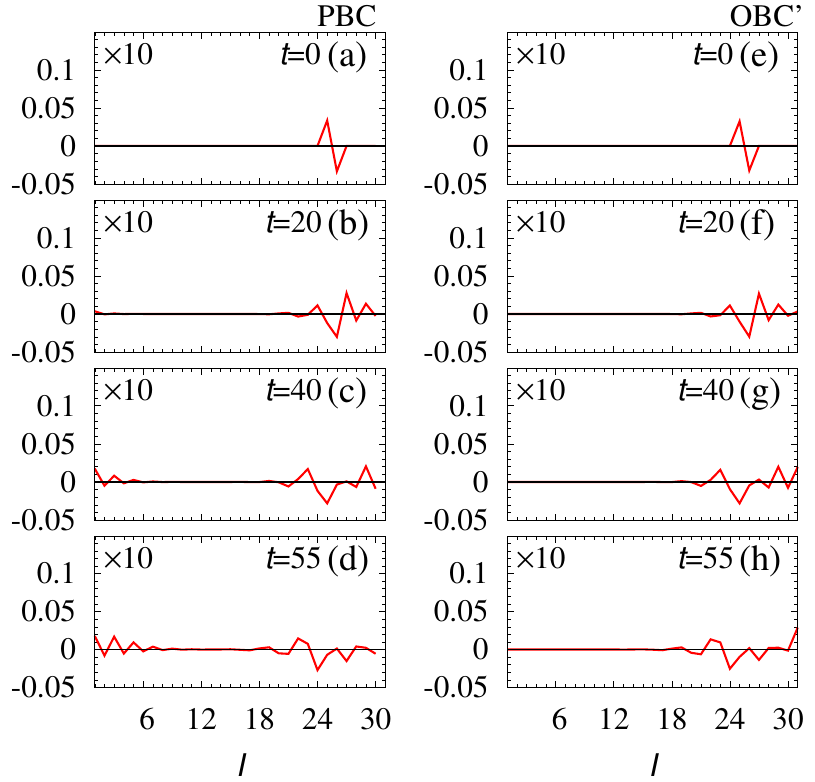}
\end{center}
\end{minipage}
\caption{
Time-evolution of the population density $\delta \bm{x}(t)=\bm{x}(t)-\bm{c}$ for $\lambda=0$ and $L_x=15$.
The horizontal axis denotes $I$. (a)-(d) [(e)-(f)]: The time-evolution under the PBC [OBC']. 
The data in all panels are multiplied by 10. 
These data are obtained in the same way as those of Fig.~4.
}
\label{fig: HermiK dyn}
\end{figure}

Figure~\ref{fig: HermiK dyn} displays the dynamical properties for $\lambda=0$ where the payoff matrix is anti-Hermitian.
Because all eigenvalues $\epsilon$'s are pure imaginary, the directive perpetration which arises from the real-part of the eigenvalues $\epsilon$'s is not observed.
In addition, no localized mode is enhanced due to the absence of skin modes.

\section{
Another RPS chain
}
\label{sec: another RPS chain app}

\begin{figure}[!h]
\begin{minipage}{0.6\hsize}
\begin{center}
\includegraphics[width=1\hsize,clip]{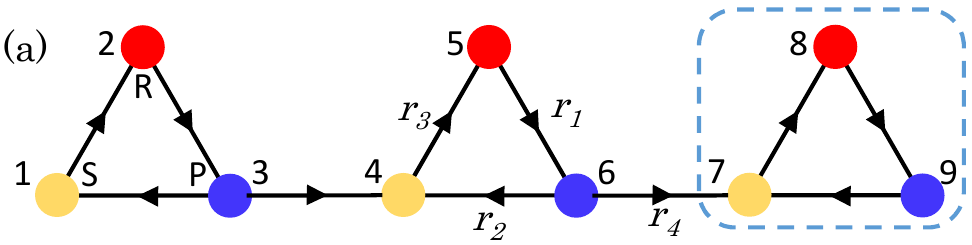}
\end{center}
\end{minipage}
\begin{minipage}{1\hsize}
\begin{center}
\includegraphics[width=1\hsize,clip]{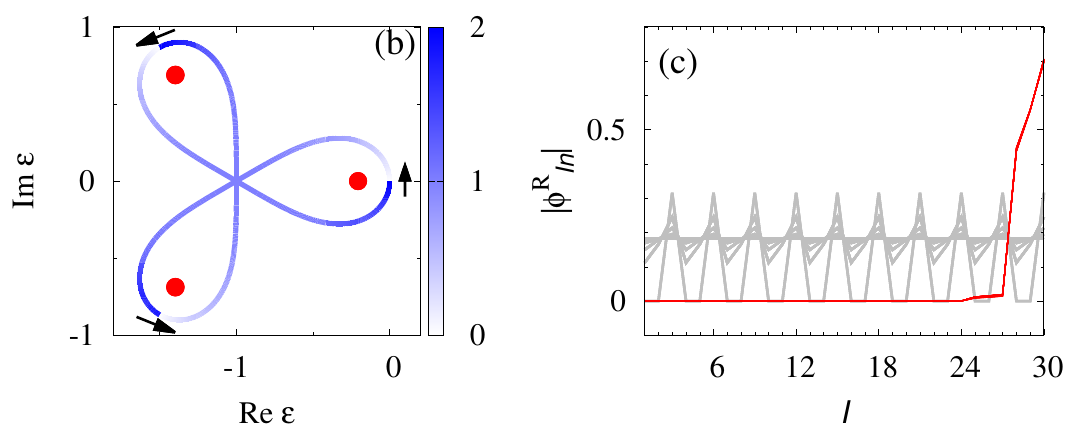}
\end{center}
\end{minipage}
\caption{
(a): Sketch of the RPS chain for $L_x=3$. 
The arrows and $r$'s describe payoffs. Here, $A_{II}=d$ for an arbitrary $I$.
As shown in this panel, $I$ takes $I=1,2,\ldots,3L_x$. For the PBC, $I+3L_x=I$ holds.
Dashed line denotes the unit cell. For $J=1,2,\ldots,L_x$, $R_{3J-2}=R_{3J-1}=R_{3J}=J$ holds.
(b):
Spectrum of the RPS chain for $(r_1,r_2,r_3,r_4,d)=(1,1/2,1,1/2,-1)$.
The data colored with blue are obtained under the PBC. Here, the shade of color denotes $k/\pi$.
The data colored with red are obtained for $L_x=10$ and the OBC.
(c): Amplitude of the right eigenvectors $|\phi^{\mathrm{R}}_{In}|$ $n=1,2,\ldots,3L_x$ as functions of $I$ for $L_x=10$.
Red (gray) lines denote data for the OBC (PBC).
For the OBC, all of the eigenvectors are localized around the right edge.
}
\label{fig: Skin spec another app}
\end{figure}

We introduce another RPS chain [see Fig.~\ref{fig: Skin spec another app}(a)] exhibiting the skin effect. 
For $L_x=5$, the payoff matrix is written as
\begin{subequations}
\begin{eqnarray}
A&=& 
\left(
\begin{array}{ccccc}
A_0          & 0           & 0           & 0           &A'_\mathrm{c} \\
A_\mathrm{c} & A_0         & 0           & 0           & 0  \\
0            &A_\mathrm{c} & A_0         & 0           & 0  \\
0            &  0          &A_\mathrm{c} & A_0         & 0  \\
0            &  0          & 0           &A_\mathrm{c} & A_0         
\end{array}
\right),
\end{eqnarray}
with
\begin{eqnarray}
A_0&=& 
\left(
\begin{array}{ccc}
d   & 0  & r_2 \\
r_3 &d   & 0 \\
 0  &r_1 & d   
\end{array}
\right),
\\
A_{\mathrm{c}}&=& 
\left(
\begin{array}{ccc}
 0  & 0  & r_4 \\
 0  & 0  & 0 \\
 0  & 0  & 0     
\end{array}
\right),
\end{eqnarray}
and the vector $\bm{x}$ consists of fifteen components,
\begin{eqnarray}
\bm{x}&=&(x_1,x_2,x_3,\cdots,x_{15})^T.
\end{eqnarray}
\end{subequations}
Here, $A'_\mathrm{c}$ is equal to $A_\mathrm{c}$ (the zero matrix) under the PBC (OBC).

Applying the Fourier transformation, the payoff matrix is written as
\begin{eqnarray}
A(k)&=& 
\left(
\begin{array}{ccc}
d   &  0  & r_2+r_4e^{ik} \\
r_3 & d   & 0 \\
0   & r_1 & d
\end{array}
\right),
\end{eqnarray}
under the PBC.
Here, $\bm{x}_{k}$ is defined as $\bm{x}^T_{k}=(x_{k\mathrm{S}},x_{k\mathrm{R}},x_{k\mathrm{P}})$ with  $x_{ks}=\frac{1}{L_x} \sum_{R_I} e^{ik R_I} x_{R_Is} $ and $s=\mathrm{R},\mathrm{P},\mathrm{S}$.
Sets of $R_I$ and $s_{I}$ are specified by $I$ ($x_I=x_{R_Is_I}$).

Figure~\ref{fig: Skin spec another app}(b) plots the spectrum of the payoff matrix for $(r_1, r_2,r_3,r_4,d)=(1,1/2,1,1/2,-1)$.
When the PBC is imposed, eigenvalues form a loop structure as denoted by blue lines in Fig.~\ref{fig: Skin spec another app}(b).
Accordingly, the winding number [Eq.~(8)] takes $1$ for $\epsilon_{\mathrm{ref}}=0.5$, which implies the skin effect.
Indeed, imposing the OBC significantly changes the spectrum [see red dots in Fig.~\ref{fig: Skin spec another app}(b)].
Correspondingly, all of the right eigenvectors are localized around the edges, meaning the emergence of skin modes [see Fig.~\ref{fig: Skin spec another app}(c)].
The above data [Figs.~\ref{fig: Skin spec another app}(b)~and~\ref{fig: Skin spec another app}(c)] indicate that the skin effect is observed in this RPS chain as well. 
The essentially same results of eigenvalues and eigenvectors are also obtained for $(r_1, r_2,r_3,r_4,d)=(2,2,2,2,1)$. 
For this parameter set, the chain is composed of an ordinary RPS cycles~\cite{Wang_RPSstandard_SciRep14} whose payoff matrix is written as
$
\left(
\begin{array}{ccc}
1 & 0& 2 \\
2 & 1& 0 \\
0 & 2& 1
\end{array}
\right)
$.
%
\begin{figure}[!h]
\begin{minipage}{0.6\hsize}
\begin{center}
\includegraphics[width=1\hsize,clip]{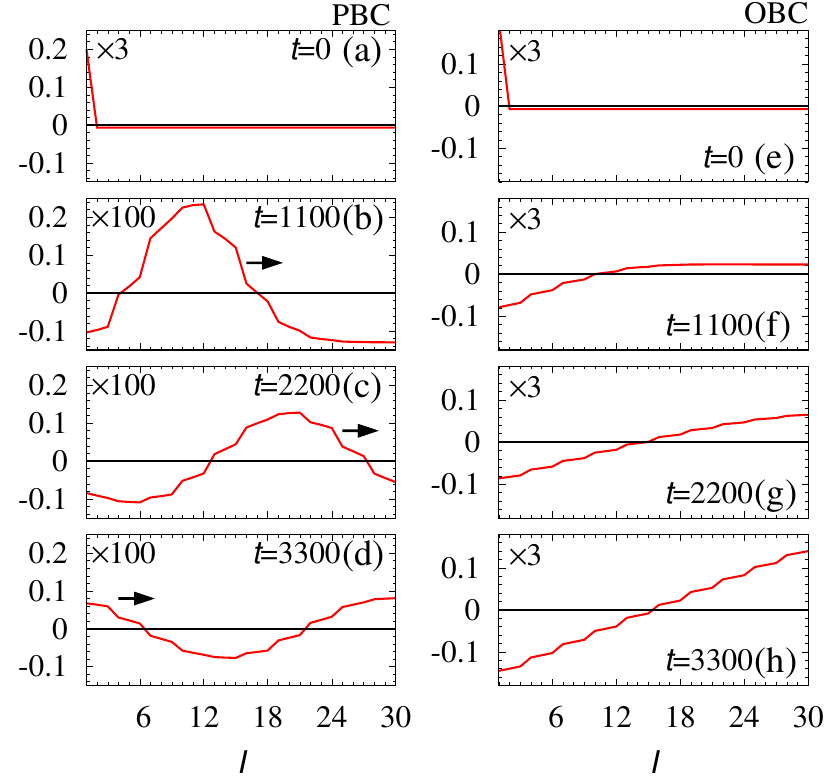}
\end{center}
\end{minipage}
\caption{
The time-evolution of the population density $\delta \bm{x}(t)=\bm{x}(t)-\bm{c}$ for $(r_1,r_2,r_3,r_4,d)=(1,1/2,1,1/2,-1)$ and $L_x=10$.
The horizontal axis denotes $I$.
(a)-(d) [(e)-(f)]: The time-evolution under the PBC [OBC]. 
The data in panels (b)-(d) [(a) and (e)-(h)] are multiplied by $100$ [$3$].
The data are obtained by employing the fourth order Runge-Kutta method~\cite{Suli_RK_textbook03} with discretized time $t_n=n\Delta t_{\mathrm{RK}}$ with $\Delta t_{\mathrm{RK}}=0.05$.
We set the initial condition as $\delta \bm{x}(0)=\frac{\delta_0}{(3L_x-1)3L_x}(3L_x-1,-1,-1,\ldots,-1)$ with $\delta_0=2$.
}
\label{fig: Skin dyn}
\end{figure}

As discussed in main text, the non-Hermitian topology characterized by $W=1$ induces the directive propagation under the PBC [see Fig.~\ref{fig: Skin dyn}(a)-\ref{fig: Skin dyn}(d)] where Eq.~(6) is satisfied for this parameter set.
We also note that under the OBC, the players prefer to be localized around the right edge for large $t$ [see Fig.~\ref{fig: Skin dyn}(e)-\ref{fig: Skin dyn}(f)]. 
However, introducing the edges violates Eq.~(6) which prevents us from mapping Eq.~(4) to the Schr\"odinger equation; introducing perturbations around the edge does not help to recover Eq.~(6) for this model.


\end{document}